\documentclass{evn2002}
\usepackage{graphicx}
\begin{document}
   \title{Is 0716+714 a superluminal blazar?}

   \author{U. Bach\inst{1}, T.P. Krichbaum\inst{1}, E. Ros\inst{1}, S. Britzen\inst{2}, A. Witzel\inst{1} and J.A. Zensus\inst{1}
          }
\authorrunning{U. Bach et al.} 
   \institute{Max Planck Institut f\"ur Radioastronomie, Auf dem H\"ugel 69, 53121 Bonn, Germany
         \and
             Max Planck Institut f\"ur Astronomie, K\"onigstuhl 17, 69117 Heidelberg, Germany
             }
   \abstract{We present an analysis of new and old high frequency
VLBI data collected during the last 10 years at 5--22\,GHz. For the jet
components in the mas-VLBI jet, two component identifications are
possible. One of them with quasi-stationary components oscillating about their
mean positions. Another identification scheme, which formally gives
the better expansion fit, yields motion with $\sim 9$\,$c$ for
$H_0=65$\,km\,s$^{-1}$\,Mpc$^{-1}$ and $q_0=0.5$. This model would be in
better agreement with the observed rapid IDV and the expected high
Lorentz-factor, deduced from IDV.
   }
   \maketitle
%
%
\section{Introduction}
The S5 blazar 0716+714 is one of the most active BL Lac objects. It is extremely variable on time scales from hours to months. 0716+714 has yet no known redshift. Optical imaging of the underlying galaxy however reveals an estimate of $z \geq 0.3$ (Wagner et al. 1996). In the radio bands 0716+71 is an intraday variable (IDV). It exhibits a very flat radio spectrum, extending up to at least 300\,GHz. The variability appears to be correlated over wide ranges of the electromagnetic spectrum (Quirrenbach et al. 1991). The simultaneous variations between X-ray, optical and radio strongly suggest an intrinsic origin of variability. From IDV a high intrinsic Doppler factor could be expected. VLBI studies covering more than 20\,years at 5\,GHz show a core-dominated evolving jet extending to the north. The VLBI jet is oriented at 90$^\circ$ with respect to the VLA jet. At present it is not clear if 0716+714 is a superluminal source and how fast the motion of the VLBI components is. There are several possible models proposed with motions between $0.05-0.5$\,mas\,yr$^{-1}$ but some of them are based on relative unfrequently sampled data (Eckart et al. 1986, 1987; Witzel et al. 1988; Schalinski et al. 1992; Gabuzda et al. 1998; Tian et al. 2000).

We present and discuss our preliminary results of the reanalysis of 10 years (25 epochs) of VLBI data on 0716+714 at frequencies between 5\,GHz and 22\,GHz. Our analysis is based on four epochs from the CJF-Survey at 5\,GHz between 1992 and 1999 (\cite{britzen}), five epochs at 8.4\,GHz from astrometric observations (\cite{ros}) and our own observations between 1994 and 1999, five epochs at 15\,GHz from the 2cm-Survey from 1994 to 2001 (\cite{Kellermann}) and 11 epochs at 22\,GHz from Jorstad et al. (2001) and our own data from 1992 to 1997.
\section{Observations and results}
All the data were available as calibrated {\sc AIPS} FITS files. The imaging of the source and the phase and amplitude self-calibration were done in {\sc Difmap}. 
   \begin{figure}
   \centering
   \includegraphics[angle=0,width=0.42\textwidth]{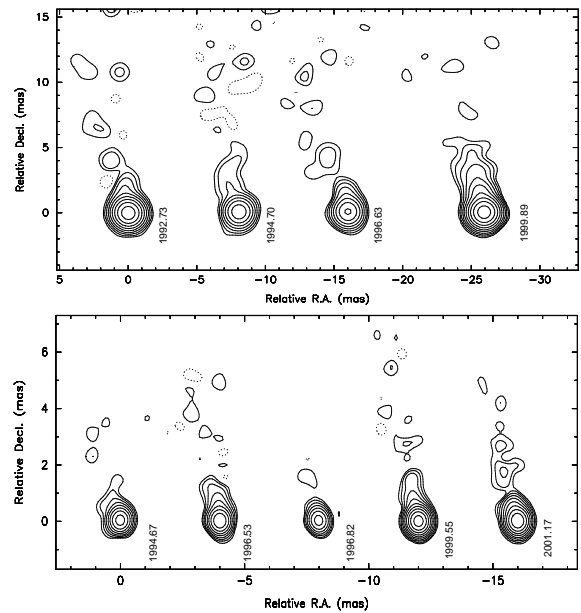}
      \caption{{\bf Top:} Four epochs of 5\,GHz maps. The maps are convolved with a 1.2\,mas circular beam respectively. The lowest contours are of 2.5, 0.9, 2.8, 1.5 and 1.1\,mJy/beam. {\bf Bottom}: Five epochs 15\,GHz maps. The maps are convolved with a 0.5\,mas circular beam respectively. The lowest contours are of 1.5, 2.5, 1.3 and 0.7\,mJy/beam.
              }
         \label{maps}
   \end{figure}
In Figure~\ref{maps}, we present a time sequence from the CJF-Survey and the 2cm-Survey maps. The jet is clearly visible to the north at P.A. $\sim 15^\circ$, but does not looks very straight. The ejection angle at the jet base shows regular variations and also the angle of the jet on larger scales seems to vary with time.
\subsection{Model-fits}
\begin{figure}
   \centering
   \includegraphics[angle=0,width=0.42\textwidth]{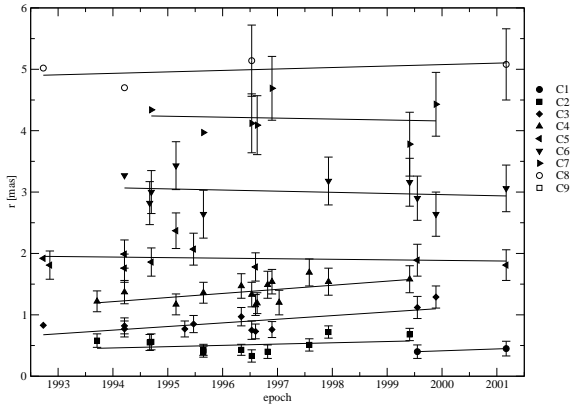}
      \caption{Trajectories of model components in the quasi-stationary model. The lines are fits to illustrate the motion of identified components.}         \label{model1}
   \end{figure}

\begin{figure}
   \centering
   \includegraphics[angle=0,width=0.42\textwidth]{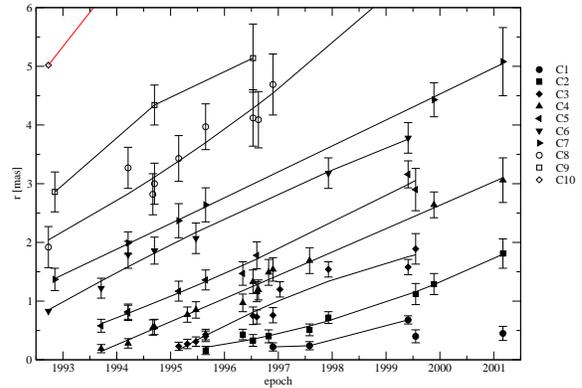}
      \caption{Trajectories of model components in the superluminal motion scenario. The lines are fits to illustrate the motion of identified components.}
         \label{model2}
   \end{figure}

After imaging in {\sc Difmap} we model fitted the self-calibrated data by circular gaussian components. We used circular components in order to reduce the degrees of freedom and for simplify the analysis. In Figures~\ref{model1}~\&~\ref{model2}, the core distance of individual VLBI components derived from our model-fits at various frequencies are plotted against time. A maximum error of 15\% in the flux density can be estimated from the amplitude calibration and the uncertainties in the model-fits. The  position error given in Figures~\ref{model1}~\&~\ref{model2} is a superposition of the map cell size and an error of 10\% of the core distance.
\section{Discussion}
\subsection{Identification of the components}
To investigate the kinematics of 0716+714 we tried to identify individual model components along the jet by their properties, e.g. distance from the core, flux density and size. Since the observations were not phase-referenced, the absolute position information is lost, and it is impossible to tell which component, if any, is stationary. Although we presently do not know if the brightest component is stationary we measured the positions with respect to the map phase center. But by a graphical analysis, which is presented in Figure~\ref{model1}~\&~\ref{model2}, we could obtain two satisfactory scenarios for the kinematics of the jet.
\subsection{Scenarios}
{\bf Model 1:} The first interpretation yields quasi-stationary components oscillating about their mean positions (see Figure~\ref{model1}). This plot gives a reasonable argument presuming that the position of the brightest components is stable. If the core position were not stable all components at one epoch should have an offset with respect to the surrounding observations which is not visible.\\
{\bf Model 2:} This model agrees with the theory that one can expect high Doppler factors for IDV sources. The components depart with $0.1$\,mas\,yr$^{-1}$, for the inner jet, and with up to $0.6$\,mas\,yr$^{-1}$, for the outer regions, from the core. Especially between 1993 and 1996, where we have many measurements spaced by only a few month, this model gives the most reasonable fit to the data. Assuming $z=0.3$ these motions would correspond to a $\beta_{\rm app}$ of 2--11 ($H_0=65$\,km\,s$^{-1}$\,Mpc$^{-1}$ and $q_0=0.5$).
\section{Conclusion}
At present we prefer Model\,2 which would be in better agreement with the observed rapid IDV. Due to the observed radio optical correlation, the rapid IDV in 0716+714 cannot be explained by refractive interstellar scattering. Therefore the source has to be very small ($T_{\rm B}\geq10^{17}$\,K) and the Doppler factor has to be $\geq50$. The large oscillation of the components in Model\,1 should make one suspicious about the identification. Further analysis with a comprehensive error calculation will show which model will prevail.
\begin{acknowledgements}
We thank S. Jorstad and A. Marscher, the group of the 2cm Survey and the group of the CJF-Survey for providing their data. U.B. and S.B. acknowledge partial support from the EC ICN RadioNET (Contract No. HPRI-CT1999-40003).
 \end{acknowledgements}

\end{document}